\documentstyle[aps,prl,preprint,floats,epsfig]{revtex}

\begin{document}
\newcommand{\mev   }{\mbox{\rm MeV}}
\newcommand{\mevc  }{\mbox{\rm MeV/$c$}}
\newcommand{\mevcsq}{\mbox{\rm MeV/$c^2$}}
\newcommand{\gev   }{\mbox{\rm GeV}}
\newcommand{\gevc  }{\mbox{\rm GeV/$c$}}
\newcommand{\gevcsq}{\mbox{\rm GeV/$c^2$}}
\newcommand{\omegac}{\mbox{$\Omega_c^0$}}
\newcommand{\casz}{\mbox{$\Xi^0$}}
\newcommand{\casmi}{\mbox{$\Xi^-$}}
\newcommand{\cascz}{\mbox{$\Xi^0_c$}}
\newcommand{\cascpl}{\mbox{$\Xi^+_c$}}
\newcommand{\kmi}{\mbox{$K^-$}}
\newcommand{\kpl}{\mbox{$K^+$}}
\newcommand{\pim}{\mbox{$\pi^-$}}
\newcommand{\pip}{\mbox{$\pi^+$}}
\newcommand{\piz}{\mbox{$\pi^0$}}
\newcommand{\kz}{\mbox{$K^0$}}
\newcommand{\ksh}{\mbox{$K^0$}}
\newcommand{\kzsh}{\mbox{$K^0_{S}$}}
\newcommand{\lz}{\mbox{$\Lambda$}}
\newcommand{\sigpl}{\mbox{$\Sigma^+$}}
\newcommand{\lcpl}{\mbox{$\Lambda^+_c$}}
\newcommand{\omgmi}{\mbox{$\Omega^-$}}
\newcommand{\dpl}{\mbox{$D^+$}}
\newcommand{\kzb}{\mbox{$\overline{K^0}$}}
\newcommand{\dz}{\mbox{$D^0$}}
\newcommand{\uq}{\mbox{$u$}}
\newcommand{\dq}{\mbox{$d$}}
\newcommand{\sq}{\mbox{$s$}}
\newcommand{\cq}{\mbox{$c$}}
\newcommand{\ubar}{\mbox{$\overline{u}$}}
\newcommand{\dbar}{\mbox{$\overline{d}$}}
\newcommand{\sbar}{\mbox{$\overline{s}$}}
\newcommand{\cbar}{\mbox{$\overline{c}$}}
\newcommand{\phot}{\mbox{$\gamma$}}
\newcommand{\decays}{\mbox{$\rightarrow$}}
\newcommand{\xp}{\mbox{$x_p$}} 
     \def\etal{{\em et al.}}

\preprint{\tighten\vbox{\hbox{\hfil CLNS 00-1695}
                        \hbox{\hfil CLEO 00-19}
}}

\title{Observation of the \omegac\ Charmed Baryon at CLEO}

\author{CLEO Collaboration}
\date{\today}

\maketitle
\tighten

\begin{abstract}

The CLEO experiment at the CESR collider has used 13.7 fb$^{-1}$ of data
to search for the production of the
\omegac\ (css-ground state) in $e^{+}e^{-}$ collisions at $\sqrt{s} \simeq 10.6$ \gev.
The modes used to study the \omegac\ are \omgmi\pip, \omgmi\pip\piz, 
\casmi\kmi\pip\pip, \casz\kmi\pip, and \omgmi\pip\pip\pim. 
We observe a signal of 40.4$\pm$9.0(stat) events
at a mass of 2694.6$\pm$2.6(stat)$\pm$1.9(syst) \mevcsq, for all modes combined.
\end{abstract}
\newpage

{
\renewcommand{\thefootnote}{\fnsymbol{footnote}}

\begin{center}
D.~Cronin-Hennessy,$^{1}$ A.L.~Lyon,$^{1}$ E.~H.~Thorndike,$^{1}$
V.~Savinov,$^{2}$
T.~E.~Coan,$^{3}$ V.~Fadeyev,$^{3}$ Y.~S.~Gao,$^{3}$
Y.~Maravin,$^{3}$ I.~Narsky,$^{3}$ R.~Stroynowski,$^{3}$
% ...
\end{center}
 
\small
\begin{center}
$^{1}${University of Rochester, Rochester, New York 14627}\\
$^{2}${Stanford Linear Accelerator Center, Stanford University, Stanford,
California 94309}\\
$^{3}${Southern Methodist University, Dallas, Texas 75275}\\
% ...
\end{center}
 
\setcounter{footnote}{0}
}
\newpage

 Many experimental groups have searched for the \omegac\ in numerous decay modes:
however, their reported \omegac\ masses are only marginally
consistent with each other. The WA62 experiment~\cite{wa62}
claimed the first evidence of \omegac\ in the \casmi\kmi\pip\pip\ decay mode with a mass of
$2740.0\pm20.0$~\mevcsq. The ARGUS Collaboration\cite{arg} published an \omegac\ 
signal in the \casmi\kmi\pip\pip\ mode with a mass of $2719.0\pm7.0\pm2.5~\mevcsq$, 
based on 0.380 fb$^{-1}$ of data. This result was contradicted by CLEO, in an unpublished 
conference paper~\cite{conf} using 1.8 fb$^{-1}$ of data. Later, E687\cite{e687}
published an \omegac\ mass of $2705.9\pm3.3\pm2.3$~\mevcsq\ using the \omgmi\pip\ mode
and a mass of $2699.9\pm1.5\pm2.5$~\mevcsq\ using the higher-statistics \sigpl\kmi\kmi\pip\  mode. 
In 1995, the WA89 Collaboration\cite{wa89} reported 200 \omegac\ events in seven decay modes
with an average mass of 2707.0$\pm$1.0(stat)~\mevcsq; this result remains unpublished.
 
 The \omegac\ (\cq\{\sq\sq\}) is a 
$J^{P}$ = $\frac{1}{2}^{+}$ ground state baryon, where \{\sq\sq\} denotes the symmetric
nature of its wave function with respect to the interchange of light-quark spins.
Various theoretical models\cite{RLP,jen,sam,am} predict an \omegac\
mass in the range 2664 - 2786~\mevcsq.

 The data used in this analysis were collected with CLEO II\cite{yk} and the upgraded
CLEO II.V\cite{hill} detector operating at the Cornell Electron Storage Ring (CESR). The
data corresponds to an integrated luminosity of 13.7 fb$^{-1}$ from the $\Upsilon(4S)$
resonance and at energies in the continuum region just below. We searched for the \omegac\ 
in the five decay modes \omgmi\pip, \omgmi\pip\piz, 
\omgmi\pip\pip\pim, \casmi\kmi\pip\pip, and \casz\kmi\pip. The choice
of these five modes is based mainly on the pattern of other charmed baryon decays.
Reconstruction efficiencies and the size of the combinatorial background
were other considerations. A sixth channel, \sigpl\kmi\kmi\pip,
was also investigated because E687\cite{e687} showed a 
significant signal in this decay mode, although CLEO has rather low efficiency
for this mode.

 Charmed baryons at CESR are either produced from the secondary decays of $B$ mesons
or directly from $e^+e^-$ annihilations to \cq\cbar\ jets. We introduce \xp\ as
the scaled momentum of a \omegac\ candidate, where $\xp = p/p_{\rm max}$, and $p_{\rm max}$ =
$\sqrt{E^{2}_{\rm b} - M^{2}}$ with $E_{\rm b}$ equal to the beam energy and 
$M$ the mass of the \omegac\ candidate.
Our search is limited to \xp\ $>$ $0.5$ or \xp\ $>$ $0.6$, depending on the decay mode, 
so as to avoid the combinatorial 
background that dominates at low \xp. Charmed baryons from $B$ meson decays
are kinematically limited to \xp\ $<$ 0.4, so our search is
limited to the \omegac\ baryons produced in the $e^+e^-$ continuum. 
We implemented $p/K/\pi$ identification by defining a joint probability for 
each hypothesis, using both the specific ionization ($dE/dx$) in the
drift chamber and the time-of-flight to the scintillation counters. A charged track
is defined to be consistent with a particular particle hypothesis if the corresponding
probability is greater than $0.1\%$. We required all the charge tracks
in all the decay modes to be consistent with their respective particle hypotheses.
To further reduce the combinatorial background we also required the
momentum of daughter pions and kaons from \omegac\ to be greater than 0.2 \gevc\ 
to 0.5 \gevc\ depending on the decay mode.

 We begin the analysis by reconstructing \lz\decays $p$\pim, \casz\decays\lz\piz,
\casmi\decays\lz\pim, \omgmi\decays\lz\kmi, and \sigpl\decays$p$\piz.
Charge conjugation is implied throughout the analysis. 
The analysis procedure for reconstructing
these particles closely follows that presented elsewhere \cite{paul,jessop,jim}.
The hyperons are required to have vertices well separated from the beamspot,
with the flight distance of the secondary \lz\ greater than that of the \casz, \casmi, 
or \omgmi. We then combine these hyperons with tracks from the
primary event vertex to reconstruct \omegac\ candidates.

\begin{figure}
\unitlength 1in
\begin{picture}(6.0,7.0)(0.0,0.0)
\put(-1.3,-3.1){\psfig{file=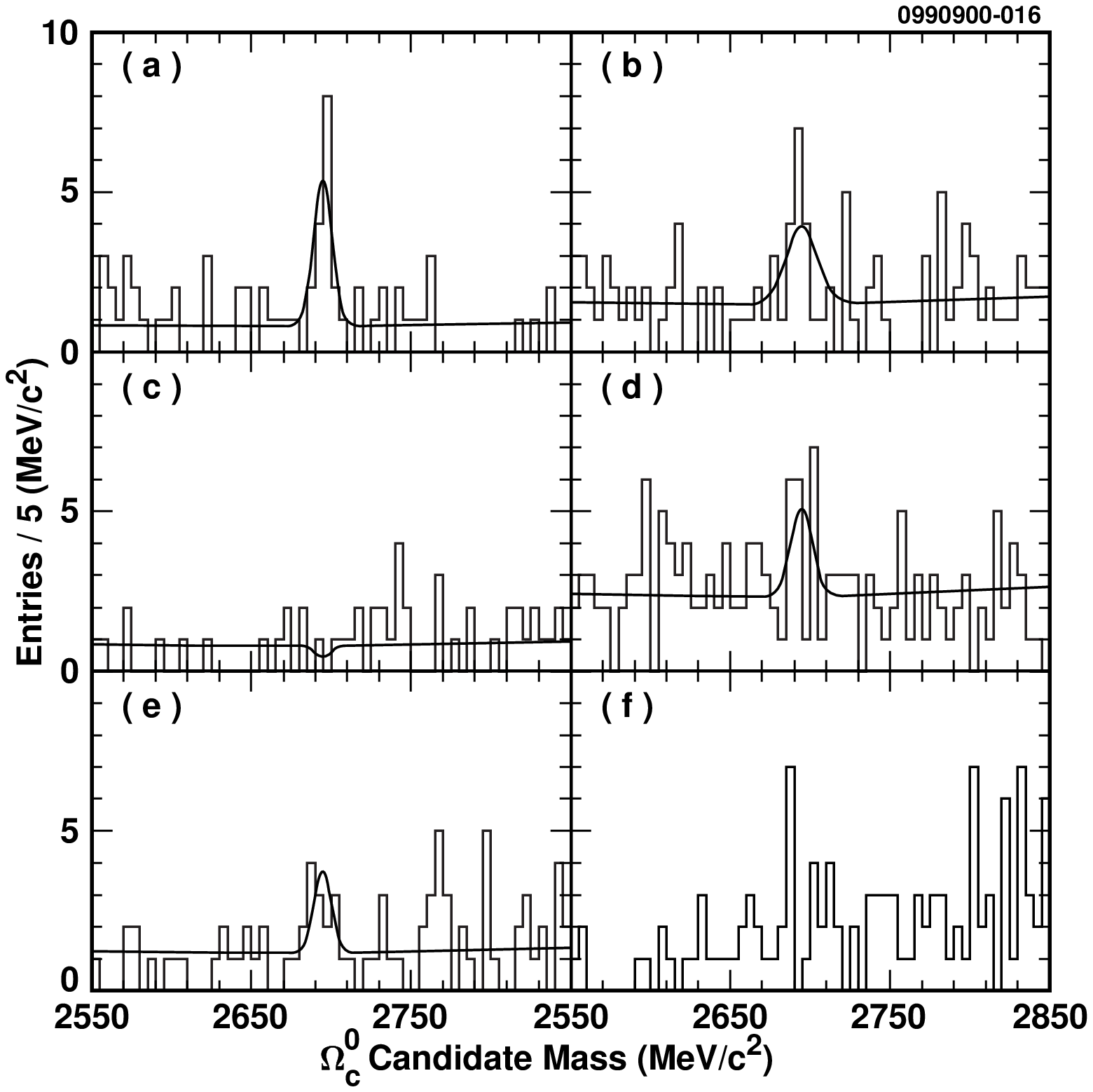,width=9.0in,bbllx=0,bblly=0,bburx=600,bbury=400}}

\end{picture}
\caption[]{
The mass distribution and simultaneous fit to the five \omegac\ modes: (a) \omgmi\pip, 
(b) \omgmi\pip\piz, (c) \omgmi\pip\pip\pim, (d) \casz\kmi\pip, (e) \casmi\kmi\pip\pip.
The mode (f) \sigpl\kmi\kmi\pip\ has not been included in the fit. The
signal is fit with a gaussian of fixed width while the background
is fit to a second order polynomial.
}
\label{fig:omczo}
\end{figure}

\begin{figure}
\unitlength 1in
\begin{picture}(6.0,7.0)(0.0,0.0)
\put(-1.3,-3.1){\psfig{file=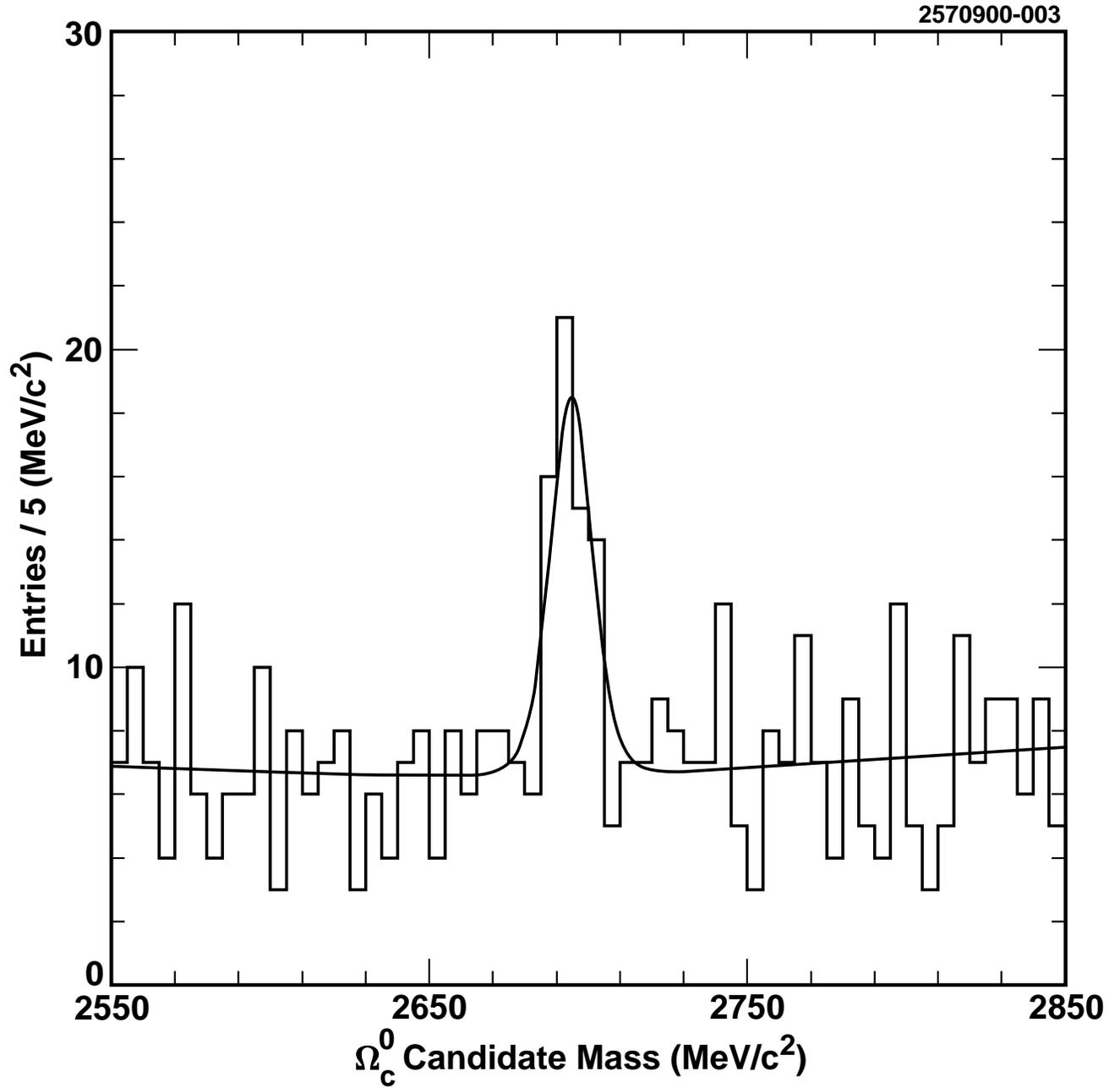,width=9.0in,bbllx=0,bblly=0,bburx=600,bbury=400}}
\end{picture}
\caption[]{
The invariant mass distribution for the sum of
\omgmi\pip, \omgmi\pip\piz, \omgmi\pip\pip\pim,  
\casz\kmi\pip, and \casmi\kmi\pip\pip combinations. The fit function
is a sum of the fit functions from Figure~\ref{fig:omczo}.

}
\label{fig:sum}
\end{figure}

In each mode the signal area above the background is obtained by fitting with the sum of a
Gaussian signal function (with width fixed at the signal Monte Carlo predicted value for that mode) 
and a second order polynomial background. The Monte Carlo sample used in this analysis was generated 
for the two CLEO detector configurations using a GEANT-based~\cite{mc} simulation and was processed 
similarly to the data. We simultaneously fit the five modes to a single mean value for the mass.
In the \omgmi\pip\ mode, we required \xp\ to be greater
than 0.5 and the \pip\ momentum to be greater than 0.5 \gevc. Figure~\ref{fig:omczo}(a)
shows the invariant mass distribution; a fit to this distribution yields a
signal of $13.3\pm4.1$ events. In the \omgmi\pip\piz\ mode, only \phot\phot\ combinations 
having invariant mass within 12.5~\mevcsq (2.5 standard deviation) of the nominal \piz\  
mass are used as \piz\ candidates; we assume the photons used for reconstructing 
\piz\decays\phot\phot\ come from the event vertex.
We also required \xp\ to be greater than 0.5 and the \pip\ and \piz\ momenta to be greater than
0.3 and 0.5 \gevc, respectively. Figure~\ref{fig:omczo}(b) shows the
invariant mass distribution. A fit to the distribution gives a signal yield of $11.8\pm4.9$ events.
Figure~\ref{fig:omczo}(c) shows the \omgmi\pip\pim\pip\ invariant mass distribution
for \xp\ greater than 0.5. All the charged pions are required
to have momenta greater than 0.2 \gevc. The corresponding fit yields a signal of $-0.9\pm1.4$ events.
In the \casz\kmi\pip\ mode, we considered combinations with \xp\ greater than 0.6,
since the combinatorial background is higher in this mode. Figure~\ref{fig:omczo}(d)
shows the invariant mass distribution, with a fit 
yielding a signal of $9.2\pm4.9$ events. In the \casmi\kmi\pip\pip\ 
mode, we required \xp\ to be greater than 0.6 and pion and kaon momenta 
to be greater than 0.2 and 0.3 \gevc, respectively. A fit to this
distribution yields a signal of  $7.0\pm3.7$ events.
Finally, in the \sigpl\kmi\kmi\pip\ mode, we required \xp\ to be greater
than 0.5 and required charged track momenta to be greater than 0.3 \gevc. 
The upper limit on the yield is 9.5 events (90 $\%$ C.L.).
Figure~\ref{fig:omczo}(f)
shows the invariant mass distribution for \sigpl\kmi\kmi\pip\ mode. 
The efficiency for \sigpl\kmi\kmi\pip\ reconstruction is $\sim 15\%$ of
that for the \omgmi\pip\ mode, which has the highest signal yield. We have not 
included the \sigpl\kmi\kmi\pip\ mode in the mass measurement. 
The total yield in the five decays modes, excluding \sigpl\kmi\kmi\pip, is $40.4\pm9.0$
events as shown in Table~\ref{tab:yield}. 
The corresponding combined mass distribution is shown in Figure~\ref{fig:sum}.

 To better determine the \omegac\ mass, we have performed an unbinned maximum-likelihood 
fit using the sum of a single 
Gaussian and a second order polynomial background. There are
two inputs to the fit, the invariant mass $M_i$ and the corresponding
mass resolution $\sigma_i$ of each of the 458 mass candidates from 2.55 to 2.85 \gevcsq. 
The invariant mass $M_i$ of each candidate is  calculated using a
vertex constrained fit; the uncertainty $\sigma_i$ is  obtained from the covariance
matrix of the fit.
The likelihood function to maximize is the product of probability density
functions (PDFs) for all the candidate events, and has the following form:

\begin{equation}
  {\cal L}(M(\omegac),f_s,a_1,a_2) = \prod_{i} \left[ f_s G(M_i - M(\omegac)|S\sigma_i)
   + (1 - f_s)\frac{P(M_i)}{\int^{2.85}_{2.55}P(M_i)dM_i} \right],
\end{equation}

 where G($y | \sigma$) = (1/$\sqrt{2\pi\sigma}$)exp($-y^2/2\sigma^2$) 
and P($y$) = $1.0 + a_1(y-2.7) + a_2(y-2.7)^2$. $M(\omegac)$ is the fitted 
\omegac\ mass, $S$ is a global scale factor multiplying $\sigma_i$, and
$f_s$ is the fraction of signal events. 
We tested the fitting procedure by first applying it to Monte
Carlo generated events with the \omegac\ mass set to 2695 \mevcsq. The fitted
mass using the above PDF is 2694.9$\pm$0.1(stat) \mevcsq\ for the
Monte Carlo and 2694.6$\pm$2.6(stat) \mevcsq\ for the data. The fitted
scale factor $S$ is 1.72$\pm$0.42 for the data and 1.16$\pm$0.02 for the
simulated events. The fitted value
of $f_s$ $=$ 0.099$\pm$0.027(stat) implies a signal size of 45.2$\pm$13.7(stat) in
good agreement with the result from Table~\ref{tab:yield}. We tested
our fitter on the \cascz\ charmed baryon in the \casmi\pip\ final state. 
Our fit gives a \cascz\ mass that is consistent with the world average value,
and the fitted global scale factor $S$ for the \cascz\ Monte Carlo and the data are in good
agreement with each other.

 We have also checked for goodness-of-fit by performing a
binned-likelihood version of the above fit (but without event-by-event mass uncertainties).
This gives an almost identical mass value and similar yields, and forms the basis for the
curve shown in Figure~\ref{fig:sum}. The $\chi^2$ for the fit to
the combined data is 46.2 for 46 degrees of freedom.

 The dominant systematic uncertainty in the mass measurement comes
from its sensitivity to the fitting method employed - the difference in the 
weighted average of the individually fitted \omegac\ modes and the unbinned
maximum likelihood method. The various fitting methods contribute 
1.5 \mevcsq\ to the systematic uncertainty. 
A systematic uncertainty (0.5 \mevcsq)
comes from imperfect treatment of modes with \piz\ mesons due to mismeasured photons
at low energies that give rise to an asymmetric \piz\ peak.
Additional contributions to the systematic uncertainty come from uncertainties
in the energy loss correction for charged tracks (0.25 \mevcsq) and the 
overall momentum scale (1.0 \mevcsq). Taking these errors
in quadrature, we estimate a total systematic uncertainty of 1.9 \mevcsq.

 In Table~\ref{tab:yield} we also give the measured inclusive cross-section times the
branching faction, $\sigma \cdot {\cal B}$, for \xp\ $>$ 0.5 into \omgmi\pip, \omgmi\pip\piz, 
\casmi\kmi\pip\pip, \casz\kmi\pip, \omgmi\pip\pip\pim and \sigpl\kmi\kmi\pip.
We estimated the systematic errors for the branching fractions by changing the
\omegac\ mass by $\pm 1.0 \sigma$ (combined error) from its best fit
value. In Table~\ref{tab:yield} the first error is due to statistics and the
second, when given, to systematics. In charm decays, $\cq\decays W\sq$, the
$W$ tends to couple more strongly to two pions (via the $\rho$-meson) than
to a single pion. The relative branching fraction 
$\cal{B}(\cq\decays\sq\pip\piz)/\cal{B}(\cq\decays\sq\pip)$ is greater
than unity. The relative branching fractions for 
$\cal{B}(\dpl\decays\kzb\pip\piz)/\cal{B}(\dpl\decays\kzb\pip)$, 
$\cal{B}(\dz\decays\kmi\pip\piz)/\cal{B}(\dz\decays\kmi\pip)$, and
$\cal{B}(\lcpl\decays\lz\pip\piz)/\cal{B}(\lcpl\decays\lz\pip)$ are
3.4$\pm$1.1(stat), 3.6$\pm$0.2(stat), and 
4.0$\pm$1.9(stat)~\cite{pdg1}, respectively. We also observe a similar
trend in the branching fraction for \omegac\decays\omgmi\pip\piz\
relative to \omegac\decays\omgmi\pip, as given in Table~\ref{tab:yield}.
We have also studied the momentum spectrum of \omegac, finding it consistent with other
charmed baryons~\cite{mom}.

\begin{table}[h]
\begin{center}
\begin{small}
\caption{ \omegac\ results in different decay modes. The fourth column shows 
the branching fractions relative to \omgmi\pip; the fifth column shows the
cross section times branching fraction. The fourth column has only statistical, while fourth
and fifth columns have statistical and systematic uncertainties.}

\label{tab:yield}
\begin{tabular}{c c c c c} \hline
\multicolumn{1}{c}{ }&$\sigma_{MC}(\mevcsq)$& Fitted Yield& Relative ${\cal B}$& $\sigma \cdot {\cal B}$ (fb) \\ 
                     &               &mode dependent \xp\ &all \xp\ $>$ 0.5&all \xp\ $>$ 0.5\\ \hline
\omgmi\pip\          & 5.87                 &13.3$\pm$4.1 &1.0             &$11.3\pm3.9\pm2.0$  \\
\omgmi\pip\piz       & 9.71                 &11.8$\pm$4.9 &$4.2\pm2.2\pm0.9$&$47.6\pm18.0\pm3.1$\\
\casz\kmi\pip        & 6.72                 &9.2$\pm$4.9  &$4.0\pm2.5\pm0.4$&$45.1\pm23.2\pm3.7$ \\
\casmi\kmi\pip\pip   & 5.46                 &7.0$\pm$3.7  &$1.6\pm1.1\pm0.4$&$18.2\pm10.6\pm3.3$\\ 
\omgmi\pip\pip\pim   & 4.89                 &-0.9$\pm$1.4 & $<$ 0.56    & $<$ 5.1  $@$ 90 $\%$ CL  \\
Combined 5 modes     &                      &40.4$\pm$9.0 &              &  \\ 
\sigpl\kmi\kmi\pip   & 6.18                 &2.8$\pm$4.1  & $<$ 4.8      & $<$ 53.8 $@$ 90 $\%$ CL \\ \hline
\end{tabular}
\end{small}
\end{center}
\end{table}               

 In conclusion, we observe the \omegac\ with a mass of 2694.6$\pm$2.6(stat)$\pm$1.9(syst) 
\mevcsq\ in the five decay modes \omgmi\pip, \omgmi\pip\piz, 
\omgmi\pip\pip\pim, \casmi\kmi\pip\pip, and \casz\kmi\pip. Although 
the signal is not statistically significant in any individual mode, the
combined signal stands out over the background with a yield of 40.4$\pm$9.0(stat)
events. Our measured $\sigma\cdot {\cal B}$ value for the 
\casmi\kmi\pip\pip\ mode ($18.2\pm10.6\pm3.3$ fb) is in clear disagreement
with the ARGUS value ($2410\pm900\pm300$ fb) for the same \xp\ range.


We gratefully acknowledge the effort of the CESR staff in providing us with
excellent luminosity and running conditions.
This work was supported by 
the National Science Foundation,
the U.S. Department of Energy,
the Research Corporation,
the Natural Sciences and Engineering Research Council of Canada, 
the A.P. Sloan Foundation, 
the Swiss National Science Foundation, 
the Texas Advanced Research Program,
and the Alexander von Humboldt Stiftung.  

\pagebreak

\clearpage
\pagebreak
\clearpage
\pagebreak
\end{document}